\documentstyle[12pt]{article}
\def\nabstar#1{\nabla\kern-0.5pt\smash{\raise 4.5pt\hbox{$\ast$}}
               \kern-4.5pt_{#1}}

\def\drvstar#1{\partial\kern-0.5pt\smash{\raise 4.5pt\hbox{$\ast$}}
               \kern-5.0pt_{#1}}

\def\newline{\relax\ifhmode\null\hfil\break\else\nonhmodeerr@\newline\fi}
\def\frac#1#2{{#1\over#2}}
\def\text#1{{\hbox{\rm #1}}}
\def\flushpar{{\par \noindent}}

\newcommand{\beq}{\begin{equation}}
\newcommand{\eeq}{\end{equation}}
\newcommand{\bea}{\begin{eqnarray}}
\newcommand{\eea}{\end{eqnarray}}
\def\Id{ \mbox{1\hspace{-1.2mm}I} }
\def\BE{\begin{equation}}
\def\EE{\end{equation}}
\def\BA{\begin{eqnarray}}
\def\EA{\end{eqnarray}}
\def\BAN{\begin{eqnarray*}}
\def\EAN{\end{eqnarray*}}
\def\LL{\left}
\def\RR{\right}
\def\nn{\nonumber\\}

\def\tr{\mbox{tr}}

\def\gm5{\gamma_5}

\def\ss#1{\slash\hspace{-2mm}#1}

\input epsf.sty
\newdimen\psfigsize
\def\psfigure#1 #2 #3 #4 #5{
    \begin{figure}[tbh]
      \begin{center}
      \vbox{
        \null\vskip-0.2in\hskip#2
        \epsfxsize=#1
        \epsfbox{#4}
        \vskip -0.3in
        \caption {#5 \label{#3}}
        \vskip 0.0 true in plus 0.3 true in
      }
      \end{center}
   \end{figure}
}
\begin{document}
\thispagestyle{empty}
\begin{flushright}
NTUTH-01-505B \\
September 2001
\end{flushright}
\bigskip\bigskip\bigskip
\vskip 2.5truecm
\begin{center}
{\LARGE {Perturbation Calculation of the Axial Anomaly of a
         Ginsparg-Wilson lattice Dirac Operator}}
\end{center}
\vskip 1.0truecm
\centerline{Ting-Wai Chiu and Tung-Han Hsieh}
\vskip5mm
\centerline{Department of Physics, National Taiwan University}
\centerline{Taipei, Taiwan 106, Taiwan.}
\centerline{\it E-mail : twchiu@phys.ntu.edu.tw}
\vskip 2cm
\bigskip \nopagebreak \begin{abstract}
\noindent

A recent proposal suggests that even if a Ginsparg-Wilson lattice Dirac
operator does not possess any topological zero modes in
topologically-nontrivial gauge backgrounds, it can
reproduce correct axial anomaly for sufficiently smooth gauge
configurations, provided that it is exponentially-local, doublers-free,
and has correct continuum behavior. In this paper,
we calculate the axial anomaly of this lattice Dirac operator in weak
coupling perturbation theory, and show that it recovers the topological
charge density in the continuum limit.



\end{abstract}
\vskip 1.5cm

\newpage\setcounter{page}1

\section{Introduction}

Recently, one of us ( TWC ) has constructed \cite{Chiu:2001bg}
a Ginsparg-Wilson Dirac operator which is $ \gamma_5 $-hermitian,
exponentially-local, doublers-free, and has correct continuum behavior,
but it does {\it not} possess any topological zero modes for
topologically-nontrivial background gauge fields. This suggests that
one might have the option to turn off the topological zero modes of a
Ginsparg-Wilson lattice Dirac operator, without affecting its physical
behaviors ( axial anomaly, fermion propagator, etc. ) at least for the
topologically-trivial gauge sector. Therefore it is interesting to verify
explicitly that it indeed reproduces the continuum axial anomaly for
smooth gauge backgrounds, in the framework of weak coupling perturbation
theory which is amenable to analytic calculations.

The Ginsparg-Wilson lattice Dirac operator proposed in
Ref. \cite{Chiu:2001bg} is
\bea
\label{eq:DcD}
D = a^{-1} D_c ( \Id + r D_c )^{-1} \ , \hspace{4mm} r > 0 \ ,
\eea
with
\bea
\label{eq:Dc}
D_c &=& \sum_{\mu} \gamma^\mu T^{\mu} \ ,
\hspace{4mm} T^\mu = f t^\mu f \ , \\
\label{eq:f}
f   &=& \left( \frac{2c}{\sqrt{t^2 + w^2} + w } \right)^{1/2} \ ,
\hspace{4mm} t^2 = -\sum_{\mu} t^\mu t^\mu \ .
\eea
Here $ \gamma^\mu t^\mu $ is the naive lattice fermion operator
and $ -w $ is the Wilson term with a negative mass $ -c $ ( $ 0 < c < 2 $ )
\bea
\label{eq:tmu}
t^\mu (x,y) &=& \frac{1}{2} [ U_{\mu}(x) \delta_{x+\hat\mu,y}
                       - U_{\mu}^{\dagger}(y) \delta_{x-\hat\mu,y} ] \ , \\
\label{eq:Umu}
U_\mu(x) &=& \exp
             \left[ i a g A_\mu \left( x+\frac{a}{2}\hat\mu \right) \right] \ ,
\eea
\bea
\gamma^\mu &=& \left( \begin{array}{cc}
                            0                &  \sigma_\mu    \\
                    \sigma_\mu^{\dagger}     &       0
                    \end{array}  \right)  \ ,
\eea
\beq
\sigma_\mu \sigma_\nu^{\dagger} + \sigma_{\nu} \sigma_\mu^{\dagger} =
2 \delta_{\mu \nu} \ ,
\eeq
\bea
\label{eq:w}
w(x,y) =  c - \frac{1}{2} \sum_\mu \left[ 2 \delta_{x,y}
                - U_{\mu}(x) \delta_{x+\hat\mu,y}
                - U_{\mu}^{\dagger}(y) \delta_{x-\hat\mu,y} \right] \ ,
\hspace{4mm} 0 < c < 2 \ ,
\eea
where the Dirac, color and flavor indices have been suppressed.
Note that in (\ref{eq:DcD}), we do not fix $ r = 1/2c $, but keep
it as a free parameter, since we intend to show that
{\it the axial anomaly is independent of} $ r $
{\it in the continuum limit}.

Evidently, the lattice Dirac operator (\ref{eq:DcD}) is
$ \gamma_5 $-hermitian
\bea
\label{eq:g5_hermit}
D^{\dagger} = \gamma_5 D \gamma_5 \ ,
\eea
and satisfies the Ginsparg-Wilson relation \cite{Ginsparg:1982bj}
\bea
\label{eq:gwr}
D \gamma_5 + \gamma_5 D = 2 r a D \gamma_5 D \ .
\eea

In the free fermion limit, $ D $ is exponentially-local, doublers-free,
and has correct continuum behavior \cite{Chiu:2001bg}.
These properties should be sufficient to guarantee that $ D $ can reproduce
correct axial anomaly in a smooth gauge background.

In the next section, we calculate the axial anomaly of (\ref{eq:DcD})
in weak coupling perturbation theory, and show that, in the continuum
limit, it recovers the topological charge density of the gauge background
\BAN
\label{eq:rho_c}
  \frac{g^2}{32\pi^2} \sum_{\mu\nu\lambda\sigma}
  \epsilon_{\mu\nu\lambda\sigma} \tr( F_{\mu\nu} F_{\lambda\sigma} ) \ ,
\EAN
where
\BAN
F_{\mu\nu} = \partial_\mu A_\nu - \partial_\nu A_\mu + i g [A_\mu, A_\nu] \ .
\EAN

Before we proceed to the perturbation calculations in the next section,
we first clarify that even though a Ginsparg-Wilson lattice Dirac
operator can reproduce the continuum axial anomaly for sufficiently smooth
gauge configurations, it does {\it not} necessarily imply that it could
possess topological zero modes for topologically-nontrivial gauge backgrounds.
This can be seen as follows.

Consider a gauge configuration with positive definite
topological charge $ Q > 0 $, and its topological charge density
$ \rho(x) $ is {\it very smooth} for all $ x $,
\bea
\label{eq:rho}
\int d^4 x \ \rho(x) = Q \ .
\eea
Then, in principle, $ \rho(x) $ can be decomposed into two parts,
\bea
\rho(x) = \rho_w(x) + \rho_0(x) \ ,
\eea
where $ \rho_w(x) $ is the "winding number density" which is
positive for all $ x $, and it is the {\it smoothest} particular solution
of (\ref{eq:rho}),
\bea
\int d^4 x \ \rho_w(x) = Q \ , \hspace{4mm} \rho_w(x) > 0 \ \forall x \ ,
\eea
and $ \rho_0(x) $ is the solution of the
homogeneous equation of (\ref{eq:rho}),
\bea
\int d^4 x \ \rho_0 (x) = 0 \ .
\eea

If one assumes that the amount of the topological charge $ Q $ is bounded by
a constant times the square root of the space-time volume, then in the
infinite volume and continuum limit, $ \rho_w(x) $ becomes infinitesimally
small at any $ x $, thus it becomes {\it unobservable}.
In other words, even if one calculates the axial anomaly $ {\cal A}(x) $
(\ref{eq:anx}) of a Ginsparg-Wilson lattice Dirac operator, and shows
that it agrees with $ \rho(x) \simeq \rho_0(x) $ in the infinite volume
and continuum limit, however, one still cannot be sure whether the
integral $ \int d^4 x {\cal A}(x) $ is equal to the topological
charge $ Q $ or not.

In order to have a definite answer to this question,
one needs to perform a {\it nonperturbative}
calculation of the axial anomaly on a {\it finite} lattice.
Otherwise, it could hardly detect the "winding number density"
$ \rho_w(x) $ at a site. However, to our knowledge, there is no preceding
{\it nonperturbative analytic} calculations of the axial anomaly for
any Ginsparg-Wilson lattice Dirac operator, on a {\it finite} lattice.
Therefore, in this paper, we do not intend to tackle the difficult
problem of performing a {\it nonperturbative analytic} calculation of the
axial anomaly on a {\it finite} lattice.

Nevertheless, from the argument presented in Ref. \cite{Chiu:2001bg},
it is clear that (\ref{eq:DcD}) does not possess any topological
zero modes. Thus, it follows that, on a finite lattice, the axial anomaly
for at least some of the lattice sites could not agree with the topological
charge density, since its sum over all sites is equal to the zero index,
rather than the nonzero topological charge $ Q $ of the gauge background.
Therefore, for any {\it topologically-nontrivial} gauge background,
we have no difficulties to understand why the axial anomaly of
(\ref{eq:DcD}) can not be exactly equal to the topological charge
density for all sites on a {\it finite} lattice.
However, in the infinite volume and continuum limit, $ \rho_w(x) \to 0 $,
then the axial anomaly $ {\cal A}(x) $ can appear to agree with the
topological charge density $ \rho(x) $ for sufficiently smooth gauge
backgrounds, even though the integral $ \int d^4 x {\cal A}(x) $ is
always zero. Thus, for a sufficiently smooth gauge background
( except for any possible "winding" at the infinity ), we can
calculate the axial anomaly ( at any point except the infinity )
of the lattice Dirac operator (\ref{eq:DcD})
in weak coupling perturbation theory, and show that in the continuum limit,
it agrees with the topological charge density of the gauge background,
regardless of the global topological charge.

\section{Perturbation Calculation of the Axial Anomaly}

The axial anomaly\footnote{It is understood that the axial
anomaly (\ref{eq:anx}) at a lattice site has to be normalized by
the volume of the unit cell, $ a^4 $, when it is compared with
the topological charge density in the continuum.}
of a flavor-singlet of Ginsparg-Wilson lattice
fermions can be written as \cite{Luscher:1998pq}
\bea
\label{eq:anx}
{\cal A} (x) &=& \tr[ \gamma_5 ( \Id - r a D ) (x,x) ] \nn
             &=& \int \frac{d^4 p}{(2\pi)^4} \int \frac{d^4 q}{(2\pi)^4}
                 \ e^{i(p-q) \cdot x } \
                 \tr[ \gamma_5 ( \Id - r a D ) (p,q) ] \ ,
\eea
where the trace runs over the Dirac, color and flavor space.
Now we expand $ D $ in power series of the gauge coupling
( through the link variables )
\bea
D = D_0 + D_1 + D_2 + D_3 + D_4 + O(g^5)
\eea
where $ D_n $ denotes the terms containing the factor $ g^n $,
the gauge coupling to the $n$-th power.
Then we have
\bea
\label{eq:D_series}
& & \hspace{-4mm} \Id - r a D \nn
&=& \hspace{-4mm} \ e_0 - e_0 d_1 e_0 -  e_0 ( d_2 - d_1 e_0 d_1 ) e_0
    - e_0 ( d_3 - d_1 e_0 d_2 - d_2 e_0 d_1 + d_1 e_0 d_1 e_0 d_1 ) e_0 \nn
& & - e_0 ( d_4 - d_1 e_0 d_3 - d_3 e_0 d_1 - d_2 e_0 d_2
            + d_1 e_0 d_1 e_0 d_2 + d_1 e_0 d_2 e_0 d_1 + \nn
& & \hspace{8mm} + d_2 e_0 d_1 e_0 d_1
                 - d_1 e_0 d_1 e_0 d_1 e_0 d_1 ) e_0 + O(g^5)
\eea
where
\bea
\label{eq:d0}
e_0 &=& \Id - r a D_0 \equiv ( \Id + d_0 )^{-1} \ , \\
\label{eq:d1}
d_1 &=& r \sum_\mu \gamma_\mu ( f_1 t^\mu_0 f_0 +
                                  f_0 t^\mu_1 f_0 +
                                  f_0 t^\mu_0 f_1    ) \ , \\
\label{eq:d2}
d_2 &=& r \sum_\mu \gamma_\mu ( f_2 t^\mu_0 f_0 +
                                         f_0 t^\mu_2 f_0 +
                                         f_0 t^\mu_0 f_2 +
                                         f_1 t^\mu_1 f_0 +
                                         f_1 t^\mu_0 f_1 +
                                         f_0 t^\mu_1 f_1    ) \ , \\
\label{eq:d3}
d_3 &=& r \sum_{\mu} \gamma_{\mu}
          \sum_{i=0}^3 \sum_{j=0}^{3-i} (f_i t^{\mu}_j f_{3-i-j}) \ ,  \\
\label{eq:d4}
d_4 &=& r \sum_{\mu} \gamma_{\mu}
          \sum_{i=0}^4 \sum_{j=0}^{4-i} (f_i t^{\mu}_j f_{4-i-j}) \ , \\
\label{eq:f_series}
f &=& f_0 + f_1 + f_2 + f_3 + f_4 + O(g^5) \ , \\
\label{eq:tmu_series}
t^\mu &=& t^\mu_0 + t^\mu_1 + t^\mu_2 + t^\mu_3 + t^\mu_4 + O(g^5) \ .
\eea

Evidently some terms in (\ref{eq:D_series}) do not contribute to the
axial anomaly, since they only have a factor of a product
less than four distinct $ \gamma $ matrices, thus they vanish after
taking the trace with $ \gamma_5 $ in (\ref{eq:anx}).
Note that each term of $ d_n $ only has one factor of $ \gamma $ matrix,
and $ e_0 $ contributes at most one factor of $ \gamma $ matrix
[ see Eq. (\ref{eq:e0}) ]. Thus, the terms $ e_0 $, and $ e_0 d_n e_0 $
( for any $ n $ ) can be dropped from the series,
when (\ref{eq:D_series}) is substituted into (\ref{eq:anx}).

Then the axial anomaly at the order $ g^2 $ is due to the fourth term in
(\ref{eq:D_series}),
\bea
\label{eq:anx_2}
{\cal A}_2 (x) = \int_p \int_q \int_k
                  e^{i(p-q) \cdot x }
  \tr[ \gamma_5 e_0(p) d_1(p,k) e_0(k) d_1(k,q) e_0(q) ]
\eea
where
\BAN
\int_p \equiv \int \frac{d^4 p}{(2\pi)^4} \ , \ \mbox{etc.}
\EAN

Similarly, substituting higher order terms of (\ref{eq:D_series})
into (\ref{eq:anx}) gives the axial anomaly at orders $ g^3 $ and $ g^4 $,
respectively. Although there seems to be many terms involving
higher order vertices ( $ d_2, d_3 $, etc. ) at higher orders,
the nonvanishing contributions at orders $ g^3 $ and $ g^4 $ only
come from terms $ e_0 d_1 e_0 d_1 e_0 d_1 e_0 $
and $ e_0 d_1 e_0 d_1 e_0 d_1 e_0 d_1 e_0 $ respectively, as we
shall show later. At this point, it is instructive to identify
$ e_0 = ( \Id + d_0 )^{-1} $ (\ref{eq:d0}) with the free fermion propagator,
$ d_1 $ (\ref{eq:d1}) with the "quark-quark-gluon" vertex,
$ d_2 $ (\ref{eq:d2}) with the "quark-quark-gluon-gluon" vertex, etc.
Then the expression $ \tr( \gamma_5 e_0 d_1 e_0 d_1 e_0 ) $ contributing
to the axial anomaly at $ O(g^2) $, (\ref{eq:anx_2}), corresponds to
the triangle diagram coupling the axial current to two external gauge
bosons through an internal fermion loop.
Similarly, the expressions $ \tr( \gamma_5 e_0 d_1 e_0 d_1 e_0 d_1 e_0 ) $
and $ \tr( \gamma_5 e_0 d_1 e_0 d_1 e_0 d_1 e_0 d_1 e_0 ) $,
contributing to the axial anomaly at $ O(g^3) $ and $ O(g^4) $,
correspond to the quadrilateral ( box ) and the pentagon diagrams
coupling the axial current to three and four external gauge bosons,
respectively.
These are exactly the Feynman diagrams\footnote{See, for example,
Figure 22.2 in Ref. \cite{Weinberg:1996kr}} which contribute to the axial
anomaly in continuum nonabelian gauge theories \cite{Bardeen:1969md}.
Note that even though higher order vertices are
allowed at finite lattice spacing, the symmetry forbids their
contributions to the axial anomaly, thus only the
"quark-quark-gluon" vertex $ d_1 $ can enter the fermion loop
contributing to the axial anomaly.

In the following three subsections, we evaluate the axial anomaly at
orders $ g^2 $, $ g^3 $ and $ g^4 $ respectively.

\subsection{The Axial Anomaly at the Order $ g^2 $}

To evaluate (\ref{eq:anx_2}), first, we obtain $ d_1 $ (\ref{eq:d1}) by
expanding $ f $ and $ t^\mu $ in power series of the gauge coupling,
(\ref{eq:f_series}) and (\ref{eq:tmu_series}).

Consider
\bea
\label{eq:h}
h = \sqrt{ t^2 + w^2 } + w = h_0 + h_1 + h_2 + h_3 + h_4 + O(g^5) \ ,
\eea
which satisfies the identity $ f \cdot f \cdot h = 2 c $, i.e.,
\bea
\label{eq:ffh}
( f_0 + f_1 + \cdots ) ( f_0 + f_1 + \cdots ) ( h_0 + h_1 + \cdots ) = 2 c \ .
\eea
To order $ g $, it gives
\BAN
f_0 f_0 h_0 &=& 2 c                          \\
f_1 f_0 h_0 + f_0 f_1 h_0 + f_0 f_0 h_1 &=& 0 \ .
\EAN
Solving these two equations, we obtain
\bea
\hspace{-2mm}
f_0(p,k)
\hspace{-2mm}
&=&
\hspace{-2mm}
\left( \frac{2c}{\sqrt{ t_0^2(p)+w_0^2(p)}+w_0(p)} \right)^{1/2}
\delta^4(p-k) \equiv f_0(p) \delta^4(p-k), \\
\hspace{-2mm}
f_1(p,k)
\hspace{-2mm}
&=&
\hspace{-2mm}
-\frac{1}{2c} \frac{f_0^2(p)f_0^2(k)}{f_0(p)+f_0(k)} h_1(p,k) \ ,
\eea
where
\bea
t_0^2 (p) = -\sum_\mu t^\mu_0(p) t^\mu_0(p)
= \sum_{\mu} \sin^2 ( p_\mu a ) \ .
\eea

Similarly, we obtain
\bea
h_1(p,k)=\frac{ w_1(p,k)[w_0(p)+w_0(k)]
                - \sum_{\mu} t^{\mu}_1(p,k)[t^{\mu}_0(p) + t^{\mu}_0(k)] }
              { \sqrt{t_0^2(p)+w_0^2(p)} + \sqrt{ t_0^2(k)+w_0^2(k)} }
    + w_1(p,k) \nn
\eea

Here we recall some well-known basic formulas in
weak coupling perturbation theory ( see Appendix A ),
\BAN
t^\mu(p,k) &=& t^\mu_0(p) \delta^4(p-k) + t^\mu_1 (p,k) + O(g^2) \ , \\
w (p,k) &=& w_0 (p) \delta^4(p-k) + w_1 (p,k) + O(g^2) \ ,
\EAN
where
\BAN
t^\mu_0 (p) &=& i \sin( p_\mu a ) \ , \\
w_0 (p) &=& c - \sum_\mu [ 1 - \cos( p_\mu a ) ] \ , \\
t^\mu_1 (p,k) &=& g {\tilde A}_{\mu}(p-k)
                  \partial_\mu t^\mu_0 \left(\frac{p+k}{2} \right) \ , \\
w_1(p,k) &=& \sum_{\mu} g {\tilde A}_{\mu}(p-k)
             \partial_\mu w_0 \left( \frac{p+k}{2} \right) \ , \\
{\tilde A}_{\mu}(p-k) &=& \sum_x e^{-i (p-k) \cdot (x+\frac{a}{2} \hat\mu)}
                           A_\mu \left(x + \frac{a}{2} \hat\mu \right) \ .
\EAN

Therefore, the axial anomaly (\ref{eq:anx_2}) at the order $ g^2 $ can be
written as
\bea
\label{eq:anx_2c}
{\cal A}_2 (x) &=& g^2 \int_p \int_q \ e^{i(p-q) \cdot x } \
 \sum_{\mu,\nu} \int_k
 \tr \{ {\tilde A}_\mu (p-k) {\tilde A}_\nu (k-q) \} \times \nn
& & \tr[\gamma_5 e_0(p) d_{1,\mu}(p,k) e_0(k)
        d_{1,\nu}(k,q) e_0(q) ] \ ,
\eea
where
\bea
\label{eq:e0}
e_0(p) &=& 1 - r a D_0(p) = \frac{1}{1+d_0(p)}
  = \frac{ 1 - r f_0^2(p) \gamma \cdot t_0 (p) }
         { 1 + r^2 f_0^4(p) t_0^2 (p) } \nn
&\equiv& b(p) + \sum_\mu \gamma^\mu c^\mu (p) \ , \\
\label{eq:d0p}
d_0(p) &=&  r f_0^2(p) \gamma \cdot t_0(p) \equiv
\sum_\mu \gamma^\mu d_0^\mu (p) \ , \\
\label{eq:d1pk}
d_1(p,k) &=& g \sum_\mu {\tilde A}_{\mu}(p-k) d_{1,\mu}(p,k) \ , \\
\label{eq:d1pk_mu}
d_{1,\mu}(p,k)
&=& r \Bigl [
       f_{1,\mu}(p,k) \gamma \cdot t_0 (k) f_0(k)
        + f_0(p) \gamma \cdot t_0 (p) f_{1,\mu}(p,k) \nn
& & \hspace{8mm}
    + \ f_0(p) \gamma^\mu \partial^\mu t^{\mu}_0 (\frac{p+k}{2}) f_0(k)
        \Bigr ] \ , \\
\label{eq:f1pk}
f_{1,\mu}(p,k) &=& -\frac{1}{2c}
                     \frac{ f_0^2(p) f_0^2(k) }{ f_0(p) + f_0(k) } \times \nn
&& \hspace{-30mm}
\left[  \partial_\mu w_0 \left( \frac{p+k}{2} \right) +
\frac{  \partial_\mu w_0 \left( \frac{p+k}{2} \right) (w_0(p)+w_0(k))
    -  \partial_\mu t^\mu_0 \left(\frac{p+k}{2} \right)
     (t^{\mu}_0(p) + t^{\mu}_0(k)) }
      { \sqrt{t_0^2(p)+w_0^2(p)} + \sqrt{ t_0^2(k)+w_0^2(k)} }
 \right] \ . \nn
\eea

In the limit $ k = p $, (\ref{eq:f1pk}) and (\ref{eq:d1pk_mu}) reduce to
\bea
f_{1,\mu}(p,p) &=& \partial_{\mu} f_0(p) \ , \\
\label{eq:d1pp}
d_{1,\mu}(p,p) &=&
 \partial_{\mu} [ r f_0(p) \gamma \cdot t_0(p) f_0(p) ]
=  \partial_\mu d_0 (p) \ .
\eea

To evaluate the integral in (\ref{eq:anx_2c}), we change the
variables $ p \to p + k $ and $ q \to q + k $, then (\ref{eq:anx_2c})
becomes
\bea
\label{eq:anx_2d}
{\cal A}_2(x) =
 g^2 \int_p \int_q \ e^{i(p-q) \cdot x } \ \sum_{\mu,\nu}
  \tr \{ {\tilde A}_\mu (p) {\tilde A}_\nu (-q) \} \ G_{\mu\nu} (p,q)
\eea
where
\BAN
G_{\mu\nu}(p,q) = \int_k
\tr[\gamma_5 e_0(p+k) d_{1,\mu} (p+k,k) e_0(k)
       d_{1,\nu} (k,q+k) e_0(q+k)] \ .
\EAN

In the following, we show that
\bea
\label{eq:Gmn_2}
G_{\mu\nu}(p,q) &=& \sum_{\lambda,\sigma}
  p_\lambda q_\sigma  \frac{\partial}{\partial p_\lambda}
 \frac{\partial}{\partial q_\sigma } G_{\mu\nu}(p,q)|_{p,q=0} + O(a) \nn
&=& - M(c) \sum_{\lambda,\sigma} 4 \ p_\lambda \ q_\sigma \
                                \epsilon_{\mu\nu\lambda\sigma} + O(a)
\eea
which, when substituted into (\ref{eq:anx_2d}), leads to the axial anomaly
at the order $ g^2 $,
\bea
\label{eq:axial_anomaly}
{\cal A}_2 (x) = g^2 M(c) \sum_{\mu\nu\lambda\sigma}
                       \epsilon_{\mu\nu\lambda\sigma}
 \tr \{ ( \partial_\mu A_\nu - \partial_\nu A_\mu )
        ( \partial_\lambda A_\sigma - \partial_\sigma A_\lambda ) (x) \} \ ,
\eea
where $ M(c) $ is a coefficient which tends to $ \frac{1}{32 \pi^2} $
( for $ 0 < c < 2 $ ) in the continuum limit.

We expand $ G_{\mu\nu}(p,q) $ in power series of $ p $ and $ q $,
\bea
G_{\mu\nu}(p,q) &=& G_{\mu\nu}(0,0)
 + \sum_{\lambda}
   p_\lambda \frac{\partial}{\partial p_\lambda} G_{\mu\nu}(p,q) |_{p,q=0}
 + \sum_{\sigma}
   q_\sigma \frac{\partial}{\partial q_\sigma } G_{\mu\nu}(p,q) |_{p,q=0} \nn
 & & + \sum_{\lambda,\sigma}
        p_\lambda q_\sigma  \frac{\partial}{\partial p_\lambda}
   \frac{\partial}{\partial q_\sigma } G_{\mu\nu}(p,q)|_{p,q=0} + O(a) \ .
\label{eq:Gmn}
\eea
It is easy to see that the zeroth order and the first order terms in
(\ref{eq:Gmn}) vanish, by symmetry and the basic properties of
$ \gamma $ matrices,
\bea
\label{eq:trace_0}
&& \hspace{-14mm}
    \tr( \gamma_5 ) = \tr( \gamma_5 \gamma^\alpha ) =
    \tr( \gamma_5 \gamma^\alpha \gamma^\beta ) =
    \tr( \gamma_5 \gamma^\alpha \gamma^\beta \gamma^\sigma ) =
    \tr( \gamma_5 \gamma^\alpha \gamma^\beta
         \gamma^\lambda \gamma^\sigma \gamma^\delta ) = 0, \\
\label{eq:trace_4}
&& \hspace{-14mm}
    \tr( \gamma_5 \gamma^\alpha \gamma^\beta \gamma^\lambda \gamma^\sigma )
   = \tr(\Id) \epsilon_{\alpha\beta\lambda\sigma} \ .
\eea

Explicitly,
\bea
G_{\mu\nu}(0,0)
&=& \int_k
\label{eq:G00_a}
\tr \{ \gamma_5 \ e_0(k) \ \partial_\mu d_0(k) \ e_0(k) \
       \partial_\nu d_0(k) \ e_0(k) \} \nn
&=& \tr(\Id) \int_k
     \epsilon_{\alpha\beta\delta\sigma} \ c^\alpha (k) c^\delta (k) \
      b(k) \ \partial_\mu d_0^\beta(k) \ \partial_\nu d_0^\sigma(k)
\label{eq:G00}
\eea
where (\ref{eq:e0}), (\ref{eq:d0p}), (\ref{eq:d1pp}), 
(\ref{eq:trace_0}) and (\ref{eq:trace_4})
have been used. Note that repeated indices are summed over
in Eqs. (\ref{eq:G00}), (\ref{eq:G11_a}), (\ref{eq:G11_b}),
(\ref{eq:G22_a}) and (\ref{eq:G22_c}).
Evidently the integrand in (\ref{eq:G00}) vanishes
due to contraction of the completely antisymmetric tensor
$ \epsilon_{\alpha\beta\delta\sigma} $
with the symmetric tensor $ c^\alpha (k) c^\delta (k) $, i.e.,
\bea
\label{eq:antisym}
\sum_{\alpha\delta}  \epsilon_{\alpha\beta\delta\sigma} \
c^\alpha (k) c^\delta (k) = 0 \ .
\eea
Hence, we have
\BAN
 G_{\mu\nu}(0,0) = 0 \ .
\EAN

Next consider the first order term of the power series (\ref{eq:Gmn}),
\bea
& & \frac{\partial}{\partial p_\lambda} G_{\mu\nu}(p,q) |_{p,q=0} \nn
&=& \int_k \Bigl \{ \
\tr [ \gamma_5 \ e_0(k) \ \partial^{\lambda}_p d_{1,\mu}(p,k) |_{p=k} \
             e_0(k) \ d_{1,\nu}(k,k) \ e_0(k) ]  \nn
& & \hspace{4mm}
    + \ \tr[ \gamma_5 \ \partial_\lambda e_0(k) \ d_{1,\mu}(k,k) \
                        e_0(k) \ d_{1,\nu}(k,k) \ e_0(k) ] \ \Bigl\}
\label{eq:G11}
\eea

From (\ref{eq:d1pk_mu}), it is easy to see that
\bea
\label{eq:d11}
\partial^{\lambda}_p d_{1,\mu}(p,k) |_{p=k} =
\sum_\sigma \gamma^\sigma  h^{\sigma}_{\lambda\mu} \ ,
\eea
where each term has one factor of $\gamma$ matrix, and the explicit
expression of $ h^{\sigma}_{\lambda\mu} $ is not required for our
purpose.

Then the first term and the second term of the integrand (\ref{eq:G11})
both vanish,
\bea
\label{eq:G11_a}
& & \tr [ \gamma_5 \ e_0(k) \ \gamma \cdot h_{\lambda\mu}(k) \
          e_0(k) \ \partial_{\nu} d_0(k) \ e_0(k) ] \nn
&=& \tr(\Id) \
    \epsilon_{\alpha\beta\delta\sigma} \ c^\alpha (k) c^\delta (k) \
    b(k) \ h^\beta_{\lambda\mu}(k) \ \partial_\nu d_0^\sigma(k) = 0 \ ,
\eea
and
\bea
& & \tr [ \gamma_5 \ \partial_{\lambda} e_0(k) \
          \partial_{\mu} d_0(k) \ e_0(k) \ \partial_{\nu} d_0(k) \ e_0 ] \nn
&=&  \tr(\Id) \
 ( \epsilon_{\delta\alpha\beta\sigma} + \epsilon_{\delta\alpha\sigma\beta} ) \
          c^{\beta}(k) b(k) \
          \partial_{\lambda}c^{\delta}(k) \
          \partial_{\mu}d^{\alpha}_0(k)   \
          \partial_{\nu}d^{\sigma}_0(k)   \        \nn
& &  + \ \tr(\Id) \
         \epsilon_{\alpha\beta\sigma\delta} \
          c^{\beta}(k) c^{\delta}(k) \
          \partial_{\lambda} b(k)    \
          \partial_{\mu} d^{\alpha}_0(k) \
          \partial_{\nu}d^{\sigma}_0(k)  \nn
&=&  0 \ ,
\label{eq:G11_b}
\eea
where (\ref{eq:e0}), (\ref{eq:d0p}), (\ref{eq:d1pp}), 
(\ref{eq:d11}), (\ref{eq:trace_0}),
(\ref{eq:trace_4}) and (\ref{eq:antisym}) have been used.

Thus
\BAN
\frac{\partial}{\partial p_\lambda} G_{\mu\nu}(p,q) |_{p,q=0} = 0 \ .
\EAN

Similarly, we have
\beq
\frac{\partial}{\partial q_{\sigma}} G_{\mu\nu}(p,q)|_{p,q=0} = 0 \ .
\eeq

Now consider the second order term of the power series (\ref{eq:Gmn}),
\bea
&& \frac{\partial}{\partial p_{\lambda}} \frac{\partial}{\partial q_{\sigma}}
        G_{\mu\nu}(p,q)|_{p,q=0} \nn
&=&
 \int_k \Bigl \{ \ \tr[ \gamma_5 \ \partial_{\lambda}e_0(k) \ d_{1,\mu}(k,k) \
  e_0(k) \ \partial^{\sigma}_q d_{1,\nu}(k,q)|_{q=k} \ e_0(k) ] \nn
&& \hspace{4mm}
   + \ \tr [ \gamma_5 \ e_0(k) \
       \partial^{\lambda}_p d_{1,\mu}(p,k)|_{p=k} \ e_0(k)
       \ d_{1,\nu}(k,k) \ \partial_{\sigma}e_0(k) ]  \nn
&& \hspace{4mm}
  + \ \tr [ \gamma_5 \ e_0(k) \
      \partial^{\lambda}_p d_{1,\mu}(p,k)|_{p=k} \ e_0(k)
      \ \partial^{\sigma}_q d_{1,\nu}(k,q)|_{q=k} \ e_0(k) ] \nn
&& \hspace{4mm}
   + \ \tr [ \gamma_5 \
       \partial_{\lambda}e_0(k) \ d_{1,\mu}(k,k) \ e_0(k)
       \ d_{1,\nu}(k,k) \ \partial_{\sigma}e_0(k) ]  \Bigr \} \ .
\label{eq:G22}
\eea
Evidently, the first three terms of the integrand are zero, by symmetry.
Explicitly, the first term is equal to
\bea
& & \tr [ \gamma_5 \ \partial_{\lambda}e_0(k) \
     \partial_\mu d_0(k) \
     e_0(k) \  \gamma \cdot h_{\sigma\nu}(k) \ e_0(k) ] \nn
&=& \tr(\Id) \
 ( \epsilon_{\delta\alpha\beta\gamma} + \epsilon_{\delta\alpha\gamma\beta} ) \
    c^{\beta}(k) \ b(k) \
    \partial_{\lambda}c^{\delta}(k) \
          \partial_{\mu}d^{\alpha}_0(k)  \
          h^{\gamma}_{\sigma\nu}(k)        \nn
& &  + \ \tr(\Id) \
         \epsilon_{\alpha\beta\gamma\delta} \
                c^{\beta}(k) \ c^{\delta}(k) \
                \partial_{\lambda} b(k) \
                \partial_{\mu}d^{\alpha}_0(k) \
                h^{\gamma}_{\sigma\nu}(k)  \nn
&=&  0 \ ,
\label{eq:G22_a}
\eea
the second term is also zero since it has the same form
of the first, and the third term is
\bea
& &
    \tr [ \ \gamma_5 \ e_0(k) \ \gamma \cdot h_{\lambda\mu}(k) \
           e_0(k) \ \gamma \cdot h_{\sigma\nu}(k) \ e_0(k)  ] \nn
&=& \tr(\Id) \
    \epsilon_{\alpha\beta\delta\gamma} \ c^\alpha (k) c^\delta (k) \
      b(k) \ h^{\beta}_{\lambda\mu}(k) \ h^{\gamma}_{\sigma\nu}(k)
= 0 \ ,
\label{eq:G22_c}
\eea
where (\ref{eq:e0}), (\ref{eq:d0p}), (\ref{eq:d1pp}), 
(\ref{eq:d11}), (\ref{eq:trace_0}),
(\ref{eq:trace_4}) and (\ref{eq:antisym}) have been used.

Thus, (\ref{eq:G22}) becomes
\bea
\label{eq:G22_d}
& & \frac{\partial}{\partial p_{\lambda}} \frac{\partial}{\partial q_{\sigma}}
        G_{\mu\nu}(p,q)|_{p,q=0} \nn
&=& \int_k \tr \{
    \partial_{\lambda}[\gamma_5 e_0(k)] \ \partial_\mu d_0(k) \ e_0(k) \
    \partial_\nu d_0(k) \ \partial_{\sigma}e_0(k) \} \ .
\eea

From (\ref{eq:e0}) and (\ref{eq:d0p}), it is easy to see that $ e_0(k) $
also satisfies the Ginsparg-Wilson relation
\bea
\label{eq:gwr_e}
\gamma_5 + e_0^{-1}(k) \gamma_5 e_0(k) = 2 \gamma_5 e_0(k) \ ,
\eea
which, after differentiation with respect to $ k_\lambda $, gives
\bea
\label{eq:gwr_d}
[ \partial_\lambda e_0^{-1}(k) ] \gamma_5 e_0(k) +
e_0^{-1}(k) \gamma_5 [ \partial_\lambda e_0(k) ] =
2 \partial_\lambda [ \gamma_5 e_0(k) ] \ .
\eea
Substituting (\ref{eq:gwr_d}) and $ d_0(k) = e_0^{-1}(k) - 1 $
[ from Eq. (\ref{eq:d0}) ] into the integrand of (\ref{eq:G22_d}),
then the integrand becomes
\BAN
& & \frac{1}{2} \tr \{  \partial_\lambda e_0^{-1}(k) \ \gamma_5 \ e_0(k) \
                 \partial_\mu e_0^{-1} (k) \ e_0(k) \
                 \partial_\nu e_0^{-1} (k) \ \partial_{\sigma}e_0(k)  \} \\
& + & \frac{1}{2} \tr \{ e_0^{-1}(k) \ \gamma_5 \ \partial_\lambda e_0(k) \
                   \partial_\mu e_0^{-1} (k) \ e_0(k) \
                   \partial_\nu e_0^{-1} (k) \ \partial_{\sigma} e_0(k) \}
\EAN
which can be further reduced to
\bea
\label{eq:G22_I}
& & - \frac{1}{2}  \ \tr \{ \gamma_5 \ \partial_\mu e_0(k) \
                            \partial_\nu e_0^{-1} (k) \
                            \partial_{\sigma} e_0(k) \
                            \partial_\lambda e_0^{-1}(k)  \} \nn
& & + \frac{1}{2}  \ \tr \{ \gamma_5 \ \partial_\lambda e_0(k) \
                            \partial_\mu e_0^{-1} (k) \
                            \partial_\nu e_0 (k) \
                            \partial_{\sigma} e_0^{-1}(k)  \} \ ,
\eea
where the identities
\bea
\label{eq:eie}
e_0(k) \ \partial_\mu e_0^{-1}(k) \ e_0(k) &=& -\partial_\mu e_0(k) \\
\label{eq:iei}
e_0^{-1}(k) \ \partial_\mu e_0(k) \ e_0^{-1} (k) &=& -\partial_\mu e_0^{-1} (k)
\eea
have been used.
By symmetry, it is obvious that the contribution of the second expression
of (\ref{eq:G22_I}) is equal to that of the first one, thus
(\ref{eq:G22_d}) becomes
\bea
\hspace{-10mm}
I_{\mu\nu\lambda\sigma} & \equiv &
      \frac{\partial}{\partial p_{\lambda}}
      \frac{\partial}{\partial q_{\sigma}}  G_{\mu\nu}(p,q)|_{p,q=0} \nn
&=& \int \frac{d^4k}{(2\pi)^4} \
             \tr \{ \gamma_5 \ \partial_\mu e_0(k) \
                            \partial_\nu e_0^{-1} (k) \
                            \partial_{\lambda} e_0(k) \
                            \partial_{\sigma} e_0^{-1}(k)  \} \nn
&=& \int \frac{d^4k}{(2\pi)^4}
    \tr \left\{ \gamma_5 \partial_\mu \left( \frac{1}{1+d_0(k)} \right)
                         \partial_\nu d_0(k)
                \partial_{\lambda} \left( \frac{1}{1+d_0(k)} \right)
                \partial_{\sigma} d_0(k) \right\}
\label{eq:G22_e}
\eea
where the domain of integration is the 4-torus,
$ {\cal T}_4 = \otimes_{i=1}^4 [ -\pi/a, \pi/a ] $,
in which the endpoints ( $ \pm \pi/a $ ) in each direction are
identified to be the same point.
In the limit $ a \to 0 $, $ {\cal T}_4 \to
\otimes_{i=1}^4 \left( -\infty, \infty \right) $,
which is invariant under the transformation
\bea
\label{eq:change}
k_\mu  \to  -\frac{1}{ r^2 k_\mu } \ ,
\hspace{4mm} \mu = 1, \cdots, 4  \ ,
\eea
for any $ r \ne 0 $.

Now we change the variables according to (\ref{eq:change}), then
(\ref{eq:G22_e}) becomes
\bea
I_{\mu\nu\lambda\sigma}
= \int \frac{d^4 k}{(2\pi)^4}
    \tr \left\{ \gamma_5 \partial_\mu \left( \frac{1}{1+d_0(K)} \right)
                         \partial_\nu d_0(K)
                \partial_{\lambda} \left( \frac{1}{1+d_0(K)} \right)
                \partial_{\sigma} d_0(K) \right\}
\label{eq:G22_f}
\eea
where the arguments of $ d_0 $ are $ K_\mu = - ( r^2 k_\mu )^{-1} $.
Note that $ d k_\mu \partial_\mu $ is invariant under the
transformation (\ref{eq:change}).

Therefore, our ansatz of evaluating (\ref{eq:G22_f}) in the continuum
limit is to substitute $ d_0(K) $ by $ d_0^{-1}(k) $ in the integrand
of (\ref{eq:G22_f}), since $ d_0(k) \to i r a \ss{k} $ as $ a \to 0 $,
and the trace with $ \gamma_5 $ picks out four distinct $ \gamma $ matrices,
one from each of the four factors with partial derivatives.
Then (\ref{eq:G22_f}) reads as
\bea
\hspace{-5mm}
I_{\mu\nu\lambda\sigma} &=& \int \frac{d^4k}{(2\pi)^4}
    \tr \left\{ \gamma_5 \partial_\mu \left( \frac{d_0(k)}{1+d_0(k)} \right)
                     \partial_\nu d_0^{-1}(k)
                \partial_{\lambda} \left( \frac{d_0(k)}{1+d_0(k)} \right)
                \partial_{\sigma} d_0^{-1}(k) \right\}  \nn
&=& \int \frac{d^4k}{(2\pi)^4}
      \tr \left\{ \gamma_5 \ \partial_\mu h(k) \
                           \partial_\nu h^{-1}(k)  \
                           \partial_{\lambda} h(k) \
                           \partial_{\sigma} h^{-1}(k) \right\}  \ ,
\label{eq:G22_I1}
\eea
where
\BAN
h(k) = \frac{d_0(k)}{ 1 + d_0(k) } \ .
\EAN

The integral (\ref{eq:G22_I1}) can be evaluated by first
removing an infinitesimal ball $ B_{\epsilon} $ of radius $ \epsilon $
from the origin ( $ k = 0 $ ) of the 4-torus $ {\cal T}_4 $,
then performing the integration, and finally taking $ \epsilon \to 0 $, i.e.,
\bea
\label{eq:G22_I1a}
\hspace{-10mm}
I_{\mu\nu\lambda\sigma} &=& \frac{1}{16 \pi^4} \lim_{\epsilon \to 0}
       \int_{ {\cal T}_4 \setminus B_{\epsilon} } {d^4 k } \
       \tr \{ \gamma_5 \ \partial_\mu h(k) \ \partial_\nu h^{-1}(k) \
              \partial_{\lambda} h(k) \ \partial_{\sigma} h^{-1}(k) \} \\
\label{eq:G22_I1b}
    &=& \frac{1}{16 \pi^4} \lim_{\epsilon \to 0}
       \int_{ {\cal T}_4 \setminus B_{\epsilon} } {d^4 k } \
       \partial_\mu \tr \{ \gamma_5 \ h(k) \ \partial_\nu h^{-1}(k) \
              \partial_{\lambda} h(k) \ \partial_{\sigma} h^{-1}(k) \}
\eea
where the $ \partial_\mu $ operation in (\ref{eq:G22_I1b})
produces (\ref{eq:G22_I1a}), plus three terms which
are symmetric in $ \mu \nu $, $ \mu \lambda $, and $ \mu \sigma $,
respectively, hence neither one of these three terms contributes
to the integral.

Then according to the Gauss theorem, the volume integral over
$ {\cal T}_4 \setminus B_{\epsilon} $ can be expressed as a surface
integral on the surface $ {\cal S}_{\epsilon} $ of the ball
$ B_{\epsilon} $, provided that the integrand
is continuous in $ {\cal T}_4 \setminus B_\epsilon $.
The last condition is satisfied since $ h(k) = r a D_0(k) $ is analytic,
and $ h(k) \ne 0 $ ( free of species doublings, for $ 0 < c < 2 $ )
for any $ k \in {\cal T}_4 \setminus B_\epsilon $ \cite{Chiu:2001bg}.
Thus (\ref{eq:G22_I1b}) becomes
\bea
I_{\mu\nu\lambda\sigma} = \frac{1}{ 16 \pi^4} \lim_{\epsilon \to 0}
      \int_{S_\epsilon} {d^3 s } \ n_{\mu}
      \tr \{ \gamma_5 \ h(k) \ \partial_\nu h^{-1}(k) \
             \partial_{\lambda} h(k) \ \partial_\sigma h^{-1}(k) \}
\eea
where $ n_\mu $ is the $\mu$-th component of the outward normal vector on
the surface $ S_\epsilon $. Since $ d_0(k) \to i r a \ss{k} $
as $ k \to 0 $, we can set $ d_0(k) = i r a \ss{k} $
on the surface $ S_\epsilon $ and obtain
\BA
I_{\mu\nu\lambda\sigma} &=&  \frac{1}{16 \pi^4}
    \lim_{\epsilon \to 0} \int_{S_\epsilon}  {d^3 s} \ n_{\mu}
    \tr \Biggl \{ \gamma_5
                  \left( \frac{\ss{k}}{1+ r^2 a^2 k^2} \right)
                  \frac{\gamma_\nu}{k^2}
                  \left(\frac{\gamma_\lambda}{1+ r^2 a^2 k^2}\right)
                  \frac{\gamma_\sigma}{k^2} \Biggr \} \nn
&=&  \frac{1}{ 16 \pi^4} \ \tr(\Id) \ \epsilon_{\mu\nu\lambda\sigma} \
     \lim_{\epsilon \to 0} \int_{S_\epsilon} {d^3 s} \
     \frac{ n_\mu k_\mu}{k^4 (1 + r^2 a^2 k^2)^2}
\label{eq:G22_I1e}
\EA
where we have used (\ref{eq:trace_0}), (\ref{eq:trace_4}),
and the property
\BAN
  \int_{S_\epsilon} {d^3 s} \ n_{\mu} k_{\nu} F(k^2)
= \delta_{\mu\nu} \int_{S_\epsilon} {d^3 s} \ n_{\mu} k_{\mu} F(k^2) \ .
\EAN
Finally, the result of the integral (\ref{eq:G22_I1e}) is
\bea
I_{\mu\nu\lambda\sigma} \hspace{-2mm} &=& \hspace{-2mm}
- \frac{N_f}{4 \pi^4} \epsilon_{\mu\nu\lambda\sigma} \
        \lim_{\epsilon \to 0} \frac{1}{( 1 + r^2 a^2 \epsilon^2)^2}
        \int^{2\pi}_{0} d\phi
        \int^{\pi}_0 d\theta_2 \sin \theta_2
        \int^{\pi}_0 d\theta_1 \sin^2\theta_1 \cos^2\theta_1 \nn
\hspace{-2mm} &=& \hspace{-2mm}
  - \frac{N_f}{ 8 \pi^2} \ \epsilon_{\mu\nu\lambda\sigma} \ ,
\label{eq:const}
\eea
where $ N_f $ denotes the number of fermion flavors.
Note that (\ref{eq:const}) is {\it independent} of the parameter
$ r $ in $ D $ (\ref{eq:DcD}).
From (\ref{eq:const}) and (\ref{eq:Gmn_2}), we obtain
\bea
\label{eq:Mc}
M(c) = \frac{N_f}{32 \pi^2} \ , \hspace{4mm} 0 < c < 2 \ ,
\eea
and the axial anomaly (\ref{eq:axial_anomaly}) at the order $ g^2 $,
\bea
\label{eq:anx_g2}
{\cal A}_2 (x) =
\frac{g^2 N_f}{32 \pi^2} \sum_{\mu\nu\lambda\sigma}
\epsilon_{\mu\nu\lambda\sigma} \
\tr \{ ( \partial_\mu A_\nu - \partial_\nu A_\mu )
        ( \partial_\lambda A_\sigma - \partial_\sigma A_\lambda ) (x) \} \ ,
\eea
where
\BAN
A_\mu(x) = \int_p e^{i p \cdot x} {\tilde A}_{\mu}(p) \ , \mbox{ etc. }
\EAN

\subsection{The Axial Anomaly at the Order $ g^3 $}

Inserting the order $ g^3 $ terms of (\ref{eq:D_series})
into (\ref{eq:anx}), we get\footnote{the term $ e_0 d_3 e_0 $
has been dropped since $ \tr( \gamma_5 e_0 d_3 e_0 ) = 0 $.}
\bea
\label{eq:anx_3}
\hspace{-4mm}
{\cal A}_3(x)
= g^3 \int_p\int_q e^{i(p-q)\cdot x} \{
           J_a(p,q) + J_b(p,q) \}
 - g^3\int_p\int_r\int_q e^{i(p-q)\cdot x} J_c(p,r,q)
\eea
where
\BAN
J_a(p,q) &=&  \int_k
        \tr\bigl\{\gamma_5 e_0(p) d_1(p,k)e_0(k)d_2(k,q)e_0(q)\bigr\} \ , \\
J_b(p,q) &=&  \int_k
        \tr\bigl\{\gamma_5 e_0(p) d_2(p,k)e_0(k)d_1(k,q)e_0(q)\bigr\} \ , \\
J_c(p,r,q) &=& \int_k
	\tr\bigl\{\gamma_5 e_0(p)d_1(p,r)e_0(r)d_1(r,k)
                  e_0(k)d_1(k,q)e_0(q) \bigr\} \ ,
\EAN
and $ e_0 $ and $ d_1 $ are given in (\ref{eq:e0}) and (\ref{eq:d1pk})
respectively. It is easy to show that $ J_a $ and $ J_b $ vanish
in the continuum limit, and only $ J_c $ has nonzero contribution
to the axial anomaly at the order $ g^3 $.
To proceed, we derive $ d_2 $ (\ref{eq:d2}) by expanding $ f $ and
$ t^\mu $ in power series of the gauge coupling, (\ref{eq:f_series})
and (\ref{eq:tmu_series}), similar to the procedure of deriving $ d_1 $
in Section 2.1. First, we solve for $ f_2 $ from the higher
order equations in (\ref{eq:ffh}), then together with the formulas
of $ t^\mu_n $ (\ref{eq:tmun}) and $ w_n $ (\ref{eq:wn})
derived in the Appendix A, we obtain
\bea
\label{eq:d2pk}
d_2(p,k) = g^2 \sum_\mu {\tilde A}^{(2)}_\mu(p-k) d_{2,\mu}(p,k)
         = g^2 \sum_{\mu,\nu} {\tilde A}^{(2)}_\mu(p-k)
                                \gamma^\nu d_{2,\mu}^{\nu}(p,k) \ ,
\eea
where $ {\tilde A}^{(2)}_\mu(p-k) $ is defined in (\ref{eq:An}),
and the explicit expression of $ d_{2,\mu}^{\nu}(p,k) $ is not required
for our subsequent calculations.
In general, to any order of $ g $, we have
\bea
\label{eq:dnpk}
d_n(p,k) = g^n \sum_\mu {\tilde A}^{(n)}_\mu(p-k) d_{n,\mu}(p,k)
         = g^n \sum_{\mu,\nu} {\tilde A}^{(n)}_\mu(p-k)
                              \gamma^\nu d_{n,\mu}^{\nu}(p,k) \ ,
\eea
which can be used for higher order calculations.

Using (\ref{eq:d1pk}) and (\ref{eq:d2pk}),
and changing variables $p\rightarrow p+k$ and $q\rightarrow q+k$
in $ J_a $ and $ J_b $, while $p\rightarrow p + r + k$,
$ r \rightarrow r+k $ and $q\rightarrow q+k$ in $ J_c $,
then we can rewrite (\ref{eq:anx_3}) as
\bea
\label{eq:anx_3a}
\hspace{-4mm}
{\cal A}_3(x)
= g^3 \int_p\int_q e^{i(p-q)\cdot x} \{
           I_a(p,q) + I_b(p,q) \}
 - g^3\int_p\int_r\int_q e^{i(p+r-q)\cdot x} I_c(p,r,q)
\eea
where
\bea
I_a(p,q) &=&
        \sum_{\mu,\nu}\tr[{\tilde A}_{\mu}(p){\tilde A}^{(2)}_{\nu}(-q)]
        G^{\mu\nu}_a(p,q) \ , \\
I_b(p,q) &=&
        \sum_{\mu,\nu}\tr[{\tilde A}^{(2)}_{\mu}(p){\tilde A}_{\nu}(-q)]
        G^{\mu\nu}_b(p,q) \ , \\
I_c(p,r,q) &=&
	\sum_{\mu,\nu,\lambda}
	\tr[{\tilde A}_{\mu}(p){\tilde A}_{\nu}(r){\tilde A}_{\lambda}(-q)]
        G^{\mu\nu\lambda}_c(p,r,q) \ ,
\eea
and
\bea
\label{eq:Ga}
\hspace{-6mm}
G^{\mu\nu}_a(p,q) \hspace{-2mm} &=& \hspace{-2mm} \int_k
  \tr[\gamma_5 e_0(p+k)d_{1,\mu}(p+k,k)e_0(k)d_{2,\nu}(k,q+k)e_0(q+k)] \\
\label{eq:Gb}
\hspace{-6mm}
G^{\mu\nu}_b(p,q) \hspace{-2mm} &=& \hspace{-2mm} \int_k
  \tr[\gamma_5 e_0(p+k)d_{2,\mu}(p+k,k)e_0(k)d_{1,\nu}(k,q+k)e_0(q+k)] \\
\label{eq:Gc}
\hspace{-6mm}
G^{\mu\nu\lambda}_c(p,r,q) \hspace{-2mm} &=& \hspace{-2mm} \int_k
   \tr[\gamma_5 e_0(p+r+k)d_{1,\mu}(p+r+k,r+k)e_0(r+k) \times       \nn
& &  \hspace{8mm}  d_{1,\nu}(r+k,k) e_0(k) d_{1,\lambda}(k,q+k) e_0(q+k)] \ .
\eea

In the following, we show that $ G_a $ and $ G_b $ vanish in the
continuum limit. First, we expand (\ref{eq:Ga}) in powers
series of $p$ and $q$,
\bea
\label{eq:Ga_series}
G^{\mu\nu}_a(p,q) &=& G^{\mu\nu}_a(0,0)
 + \sum_{\sigma} p^{\sigma} \frac{\partial}{\partial p^{\sigma}}
    G^{\mu\nu}_a(p,0)\bigr|_{p=0}
 + \sum_{\sigma} q^{\sigma} \frac{\partial}{\partial q^{\sigma}}
    G^{\mu\nu}_a(0,q)\bigr|_{q=0}   \nn
& &  + O(a) \ ,
\eea
where the zeroth order and first order terms can be easily shown
to be zero, by symmetry. Explicitly,
\BAN
G^{\mu\nu}_a(0,0) = \int_k
  \tr[\gamma_5 e_0(k) d_{1,\mu}(k,k) e_0(k) d_{2,\nu}(k,k)e_0(k)] = 0 \ ,
\EAN
since the integrand vanishes identically,
\BAN
\tr[ \gamma_5 e_0(k) \partial_{\mu}e_0^{-1}(k)
         e_0(k) d_{2,\nu}(k,k) e_0(k) ]
= - \tr[\gamma_5\partial_{\mu}e_0(k) d_{2,\nu}(k,k) e_0(k)] = 0 \ ,
\EAN
where Eqs. (\ref{eq:e0}), (\ref{eq:d1pp}),
(\ref{eq:eie}) and (\ref{eq:trace_0}) have been used.

To evaluate the first order terms in (\ref{eq:Ga_series}),
we observe that\footnote{
Note that each term of $ d_1 $ (\ref{eq:d1}) and $ d_2 $ (\ref{eq:d2})
has only one factor of $ \gamma $ matrix.}
\bea
\label{eq:d1h}
 \frac{\partial}{\partial p^{\sigma}} d_{1,\mu}(p+k,k)\bigl|_{p=0}
& \equiv & \sum_\lambda \gamma^{\lambda} h^{\lambda}_{\sigma\mu}(k)
= \gamma \cdot  h_{\sigma\mu}(k) \ ,
\\
\label{eq:d2g}
 \frac{\partial}{\partial q^{\sigma}} d_{2,\nu}(k,q+k)|_{q=0} & \equiv &
\sum_\lambda \gamma^{\lambda} g^{\lambda}_{\sigma\nu}(k)
= \gamma \cdot g_{\sigma\nu}(k) \ ,
\eea
which follow from (\ref{eq:d1}), (\ref{eq:d2}), (\ref{eq:d1pk_mu})
and (\ref{eq:d2pk}), where the explicit expressions of
$ h^{\lambda}_{\sigma\mu}(k) $ and $ g^{\lambda}_{\sigma\nu}(k) $ are
not required for our subsequent calculations. Then
\BAN
& & \frac{\partial}{\partial p^{\sigma}} G^{\mu\nu}_a(p,0) \bigl|_{p=0} \nn
&=& \int_k \tr\Bigl\{\gamma_5
           [\partial_{\sigma}e_0(k) \partial_{\mu} d_0(k)
             +e_0(k) \ \gamma \cdot h_{\sigma\mu}(k)]
             e_0(k) d_{2,\nu}(k,k) e_0(k)\Bigr\} = 0 \ ,
\EAN
since its integrand is equal to
\BAN
& & \tr(\Id)\epsilon_{\alpha\beta\lambda\delta} \Bigl\{ \Bigl[
        \partial_{\sigma} b(k) \partial_{\mu} d^{\alpha}_0(k)
        +b(k) h^{\alpha}_{\sigma\mu}(k) \Bigr] c^{\beta}(k)c^{\delta}(k)
        d^{\lambda}_{2,\nu}(k)\Bigr\} \nn
&+ & \tr(\Id)(\epsilon_{\alpha\beta\lambda\delta}
        +\epsilon_{\alpha\beta\delta\lambda}) \Bigl\{ \Bigl[
	\partial_{\sigma}c^{\alpha}(k)\partial_{\mu}d^{\beta}_0(k)
        +c^{\alpha}(k)h^{\beta}_{\sigma\mu}(k) \Bigr]
        c^{\lambda}(k){d}^{\delta}_{2,\nu}(k) \Bigr\} = 0 \ ,
\EAN
where (\ref{eq:e0}), (\ref{eq:d0p}), (\ref{eq:d1h}), (\ref{eq:trace_0}),
(\ref{eq:trace_4}) and (\ref{eq:antisym}) have been used.
Similarly, we have
\BAN
& & \frac{\partial}{\partial q^{\sigma}} G^{\mu\nu}_a (0,q)\bigl|_{q=0} \nn
&=&
    \int_k \tr\Bigl\{\gamma_5 e_0(k) \partial_{\mu} d_0(k) e_0(k)
            [ \gamma \cdot g_{\sigma\nu}(k) e_0(k)
        + d_{2,\nu}(k,k) \partial_{\sigma}e_0(k)]\Bigr\} = 0 \ .
\EAN
Therefore $ G^{\mu\nu}_a (p,q) = O(a) \to 0 $ in the limit $ a \to 0 $.
By the same token, $ G^{\mu\nu}_b(p,q) = 0 $ in the continuum limit.

Next we expand $ G_c $ in power series of $ p $, $ q $ and $ r $,
\bea
\label{eq:Gc_series}
G^{\mu\nu\lambda}_c (p,r,q) = G^{\mu\nu\lambda}_c(0,0,0)
  +  \sum_{\sigma} p^{\sigma} \frac{\partial}{\partial p^{\sigma}}
       G^{\mu\nu\lambda}_c(p,0,0)\bigl|_{p=0} \hspace{12mm} \nn
+ \sum_{\sigma} r^{\sigma} \frac{\partial}{\partial r^{\sigma}}
     G^{\mu\nu\lambda}_c(0,r,0)\bigl|_{r=0}
     +  \sum_{\sigma} q^{\sigma} \frac{\partial}{\partial q^{\sigma}}
     G^{\mu\nu\lambda}_c(0,0,q)\bigl|_{q=0} \ + \ O(a) \ .
\eea
It is easy to see that the zeroth order term vanishes,
\bea
\label{eq:Gc0}
& & G^{\mu\nu\lambda}_c(0,0,0) \nn
&=& \int_k \tr[\gamma_5 e_0(k)d_{1,\mu}(k,k)e_0(k)d_{1,\nu}(k,k)e_0(k)
               d_{1,\lambda}(k,k)e_0(k)]
= 0 \ ,
\eea
since its integrand vanishes,
\BAN
& & \tr[\gamma_5 e_0(k)d_{1,\mu}(k,k)e_0(k)d_{1,\nu}(k,k)e_0(k)
         d_{1,\lambda}(k,k)e_0(k)] \nn
&=& 
    \tr[\gamma_5 e_0(k)\partial_{\mu}e_0^{-1}(k)e_0(k)
       \partial_{\nu}e_0^{-1}(k)e_0(k)\partial_{\lambda}e_0^{-1}(k)e_0(k)] \nn
&=&
    \tr[\gamma_5 \partial_{\mu}e_0(k)\partial_{\nu}e_0^{-1}(k)
         \partial_{\lambda}e_0(k) ] = 0 \ ,
\EAN
where (\ref{eq:d1pp}), (\ref{eq:eie}) and (\ref{eq:trace_0}) have been used.

Next we evaluate the first order terms
\bea
& &  \frac{\partial}{\partial p^{\sigma}}
     G^{\mu\nu\lambda}_c (p,0,0)\bigl|_{p=0} \nn
&=&
     - \int_k \tr\Bigl\{\gamma_5
      [\partial_{\sigma}e_0(k)\partial_{\mu}e_0^{-1}(k)
        +e_0(k) \gamma\cdot h_{\sigma\mu}(k) ]
        e_0(k)\partial_{\nu}e_0^{-1}(k)\partial_{\lambda}e_0(k) \Bigr\} \nn
&=&
     - \int_k \tr\Bigl\{\gamma_5
       \partial_{\sigma}e_0(k)\partial_{\mu}e_0^{-1}(k)
        e_0(k)\partial_{\nu}e_0^{-1}(k)\partial_{\lambda}e_0(k) \Bigr\} \ ,
\label{eq:Gc1p}
\eea
where (\ref{eq:d1pp}), (\ref{eq:d1h}) and (\ref{eq:eie}) have been used,
and the last equality is due to the vanishing of the second term (
containing the factor $ \gamma\cdot h_{\sigma\mu}(k) $ ) in the
integrand,
\BAN
& & \tr \Bigl\{ \gamma_5 e_0(k) \gamma \cdot h_{\sigma\mu}(k)
        e_0(k)\partial_{\nu}e_0^{-1}(k)\partial_{\lambda}e_0(k) \Bigr\} \nn
&=&
     \tr(\Id) \epsilon_{\alpha\beta\gamma\delta}
     \partial_{\lambda}b(k)c^{\alpha}(k)c^{\gamma}h_{\sigma\mu}^{\beta}(k)
     \partial_{\nu}d_0^{\delta}(k)  \nn
& & + \ \tr(\Id) ( \epsilon_{\alpha\beta\gamma\delta} +
                \epsilon_{\beta\alpha\gamma\delta} )
     b(k)c^{\alpha}(k)\partial_{\lambda}c^{\delta}(k)
     h_{\sigma\mu}^{\beta}(k)\partial_{\nu}d_0^{\gamma}(k)  \nn
&=& 0 \ .
\EAN

Comparing (\ref{eq:Gc1p}) with (\ref{eq:G22_d}),
one immediately obtains\footnote{Note that
$ \partial_{\mu}e_0^{-1}(k) = \partial_{\mu} d_0 (k) $, from (\ref{eq:e0}).}
\bea
\label{eq:Gc1_p}
 \frac{\partial}{\partial p^{\sigma}} G^{\mu\nu\lambda}_c (p,0,0)\bigl|_{p=0}
= I_{\mu\nu\lambda\sigma}
= - \frac{N_f}{8\pi^2} \epsilon_{\mu\nu\lambda\sigma} \ ,
\eea
which has been evaluated in (\ref{eq:const}).

Similarly, we obtain
\bea
\label{eq:Gc1_q}
\frac{\partial}{\partial q^{\sigma}} G^{\mu\nu\lambda}_c(0,0,q)\bigl|_{q=0}
= - I_{\mu\nu\lambda\sigma}
= \frac{N_f}{8\pi^2} \epsilon_{\mu\nu\lambda\sigma} \ .
\eea

\eject

Next we evaluate
\bea
& & \frac{\partial}{\partial r^{\sigma}}
    G^{\mu\nu\lambda}_c (0,r,0)\bigl|_{r=0} \nn
&=& \int_k
   \Bigl\{\tr[\gamma_5
   \partial_{\sigma}e_0(k)\partial_{\mu}e^{-1}_0(k)e_0(k)
   \partial_{\nu}e^{-1}_0(k)e_0(k)\partial_{\lambda}e^{-1}_0(k)e_0(k)] \nn
&& \hspace{4mm}
   +\tr[\gamma_5 e_0(k)
	\partial_{\mu}e^{-1}_0(k)\partial_{\sigma}e_0(k)
	\partial_{\nu}e^{-1}_0(k)e_0(k)\partial_{\lambda}e^{-1}_0(k)e_0(k)] \nn
&& \hspace{4mm}
   +\tr[\gamma_5 e_0(k)
	\partial_{\sigma}\partial_{\mu}d_0(k)e_0(k)
	\partial_{\nu}e^{-1}_0(k)e_0(k)\partial_{\lambda}e^{-1}_0(k)e_0(k)] \nn
&& \hspace{4mm}
\label{eq:Gc1_r}
  +\tr[\gamma_5 e_0(k)\partial_{\mu}e^{-1}_0(k)e_0(k)
      (\gamma\cdot h_{\sigma\nu}(k))e_0(k)\partial_{\lambda}e^{-1}_0(k)e_0(k)]
        \Bigr\}  \\
& = & 0 \ ,
\label{eq:Gc1_r_0}
\eea
where in the integrand of (\ref{eq:Gc1_r}), the first two terms cancel each
other, and the last two terms vanish respectively, after the
( by now familiar ) manipulations using (\ref{eq:e0}),
(\ref{eq:d1pp}), (\ref{eq:eie}), (\ref{eq:iei}),
(\ref{eq:trace_0}), (\ref{eq:trace_4}) and (\ref{eq:antisym}).

Substituting (\ref{eq:Gc0}), (\ref{eq:Gc1_p}), (\ref{eq:Gc1_q}) and
(\ref{eq:Gc1_r_0}) into (\ref{eq:Gc_series}), we obtain
\bea
  G^{\mu\nu\lambda}_c(p,r,q)
= -{N_f\over 8\pi^2} \sum_{\sigma}
     (p_{\sigma}-q_{\sigma}) \ \epsilon_{\mu\nu\lambda\sigma} + O(a) \ .
\eea

Therefore, in the continuum limit, the axial anomaly (\ref{eq:anx_3a})
at $ O(g^3) $ is
\bea
{\cal A}_3(x) &=&
     {g^3 N_f\over 8\pi^2}\sum_{\mu\nu\lambda\sigma}
	\epsilon_{\mu\nu\lambda\sigma}
	\int_p\int_r\int_q e^{i(p+r-q)\cdot x} (p_{\sigma}-q_{\sigma})
        \tr[{\tilde A}_{\mu}(p){\tilde A}_{\nu}(r){\tilde A}_{\lambda}(-q)]
        \nn
&=&
        -i \ {g^3 N_f\over 32\pi^2}\sum_{\mu\nu\lambda\sigma}
	\epsilon_{\mu\nu\lambda\sigma}
	\tr\Bigl\{(\partial_{\sigma}A_{\mu}(x)-\partial_{\mu}A_{\sigma}(x))
        [A_{\nu}(x),A_{\lambda}(x)] + \nn
&& \hspace{40mm}
        [A_{\mu}(x),A_{\nu}(x)]
	(\partial_{\sigma}A_{\lambda}(x)-\partial_{\lambda}A_{\sigma}(x))
        \Bigr\} \ ,
\label{eq:anx_g3}
\eea
where
\BAN
A_\mu(x) = \int_p e^{i p \cdot x} {\tilde A}_{\mu}(p) \ , \mbox{ etc. }
\EAN

\subsection{The Axial Anomaly up to Order $ g^4 $}

Inserting the order $ g^4 $ terms of (\ref{eq:D_series})
into (\ref{eq:anx}), we have\footnote{the term $ e_0 d_4 e_0 $
has been dropped since $ \tr( \gamma_5 e_0 d_4 e_0 ) = 0 $.}
\bea
\label{eq:anx_4}
{\cal A}_4(x) &=& g^4\int_p\int_q e^{i(p-q)\cdot x}
        \Bigl\{J_1(p,q) + J_2(p,q)+ J_3(p,q)\Bigr\} \nn
&&
	-g^4\int_p\int_r\int_q e^{i(p-q)\cdot x}
        \Bigl\{ J_4(p,r,q)+ J_5(p,r,q)+ J_6(p,r,q)\Bigr\} \nn
&&
        +g^4\int_p\int_r\int_s\int_q e^{i(p-q)\cdot x} J_7(p,r,s,q)
\eea
where
\BAN
J_1(p,q) &=& \int_k
	\tr\bigl[\gamma_5 e_0(p) d_1(p,k)e_0(k)d_3(k,q) e_0(q)\bigr] \\
J_2(p,q) &=& \int_k
	\tr\bigl[\gamma_5 e_0(p) d_2(p,k)e_0(k)d_2(k,q) e_0(q)\bigr] \\
J_3(p,q) &=& \int_k
	\tr\bigl[\gamma_5 e_0(p) d_3(p,k)e_0(k)d_1(k,q) e_0(q)\bigr] \\
J_4(p,r,q) &=& \int_k
	\tr\bigl[\gamma_5 e_0(p) d_1(p,r)e_0(r)d_1(r,k)e_0(k)d_2(k,q)
	e_0(q)\bigr] \\
J_5(p,r,q) &=& \int_k
	\tr\bigl[\gamma_5 e_0(p) d_1(p,r)e_0(r)d_2(r,k)e_0(k)d_1(k,q)
	e_0(q)\bigr] \\
J_6(p,r,q) &=& \int_k
	\tr\bigl[\gamma_5 e_0(p) d_2(p,r)e_0(r)d_1(r,k)e_0(k)d_1(k,q)
	e_0(q)\bigr] \\
J_7(p,r,s,q) \hspace{-2mm} &=& \hspace{-2mm}
        \int_k
	\tr\bigl[\gamma_5 e_0(p) d_1(p,r)e_0(r)d_1(r,s)e_0(s)d_1(s,k)
	e_0(k)d_1(k,q) e_0(q)\bigr]
\EAN
Now we change the variables as follows :  \\
(i) $p\rightarrow p+k$ and $q\rightarrow q+k$ in
    $J_1$, $J_2$, and $J_3$;      \\
(ii) $p\rightarrow p+r+k$, $r\rightarrow r+k$, and $q\rightarrow q+k$ in
    $J_4$, $J_5$, and $J_6$; \\
(iii) $p\rightarrow p+r+s+k$, $r\rightarrow r+s+k$, $s\rightarrow s+k$,
and $q\rightarrow q+k$ in $J_7$. \\

Then (\ref{eq:anx_4}) becomes
\bea
\label{eq:anx_4a}
{\cal A}_4(x) &=& g^4\int_p\int_q e^{i(p-q)\cdot x}
        \Bigl\{I_1(p,q)+I_2(p,q)+I_3(p,q)\Bigr\} \nn
&&
	-g^4\int_p\int_r\int_q e^{i(p+r-q)\cdot x}
        \Bigl\{I_4(p,r,q)+I_5(p,r,q)+I_6(p,r,q)\Bigr\} \nn
&&
        +g^4\int_p\int_r\int_s\int_q e^{i(p+r+s-q)\cdot x} I_7(p,r,s,q)
\eea
where
\BAN
I_1(p,q) &=& \sum_{\mu,\nu}
        \tr[{\tilde A}_{\mu}(p){\tilde A}_{\nu}^{(3)}(-q)]
        G^{\mu\nu}_1(p,q) \\
I_2(p,q) &=& \sum_{\mu,\nu}
        \tr[{\tilde A}_{\mu}^{(2)}(p){\tilde A}_{\nu}^{(2)}(-q)]
        G^{\mu\nu}_2(p,q) \\
I_3(p,q) &=& \sum_{\mu,\nu}
        \tr[{\tilde A}_{\mu}^{(3)}(p){\tilde A}_{\nu}(-q)]
        G^{\mu\nu}_3(p,q) \\
I_4(p,r,q) &=& \sum_{\mu,\nu,\lambda}
        \tr[{\tilde A}_{\mu}(p){\tilde A}_{\nu}(r)
        {\tilde A}_{\lambda}^{(2)}(-q)]
        G^{\mu\nu\lambda}_4(p,r,q) \\
I_5(p,r,q) &=& \sum_{\mu,\nu,\lambda}
        \tr[{\tilde A}_{\mu}(p){\tilde A}_{\nu}^{(2)}(r)
        {\tilde A}_{\lambda}(-q)]
        G^{\mu\nu\lambda}_5(p,r,q) \\
I_6(p,r,q) &=& \sum_{\mu,\nu,\lambda}
        \tr[{\tilde A}_{\mu}^{(2)}(p){\tilde A}_{\nu}(r)
        {\tilde A}_{\lambda}(-q)]
        G^{\mu\nu\lambda}_6(p,r,q) \\
I_7(p,r,s,q) &=& \sum_{\mu,\nu,\lambda,\sigma}
	\tr[{\tilde A}_{\mu}(p){\tilde A}_{\nu}(r){\tilde A}_{\lambda}(s)
        {\tilde A}_{\sigma}(-q)]G^{\mu\nu\lambda\sigma}_7(p,r,s,q)
\EAN
and
\BAN
\hspace{-4mm}
&& \hspace{-6mm}
G^{\mu\nu}_1(p,q) =
	\int_k\tr[\gamma_5 e_0(p+k)d_{1,\mu}(p+k,k)e_0(k)
	d_{3,\nu}(k,q+k)e_0(q+k)]
\\
\hspace{-4mm}
&& \hspace{-6mm}
G^{\mu\nu}_2(p,q) =
	\int_k\tr[\gamma_5 e_0(p+k)d_{2,\mu}(p+k,k)e_0(k)
	d_{2,\nu}(k,q+k)e_0(q+k)]
\\
\hspace{-4mm}
&&  \hspace{-6mm}
G^{\mu\nu}_3(p,q) =
	\int_k\tr[\gamma_5 e_0(p+k)d_{3,\mu}(p+k,k)e_0(k)
	d_{1,\nu}(k,q+k)e_0(q+k)]
\\
\hspace{-4mm}
&&  \hspace{-6mm}
G^{\mu\nu\lambda}_4(p,r,q) =
    \int_k\tr[\gamma_5 e_0(p+r+k)d_{1,\mu}(p+r+k,r+k)e_0(r+k) \times \nn
&& \hspace{30mm}
              d_{1,\nu}(r+k,k)e_0(k) d_{2,\lambda}(k,q+k)e_0(q+k)]
\\
\hspace{-4mm}
&&  \hspace{-6mm}
G^{\mu\nu\lambda}_5(p,r,q) =
     \int_k\tr[\gamma_5 e_0(p+r+k)d_{1,\mu}(p+r+k,r+k)e_0(r+k) \times \nn
&& \hspace{30mm}   d_{2,\nu}(r+k,k)e_0(k)d_{1,\lambda}(k,q+k)e_0(q+k)]
\\
\hspace{-4mm}
&&  \hspace{-6mm}
G^{\mu\nu\lambda}_6(p,r,q) =
     \int_k\tr[\gamma_5 e_0(p+r+k)d_{2,\mu}(p+r+k,r+k)e_0(r+k) \times \nn
&& \hspace{30mm}    d_{1,\nu}(r+k,k)e_0(k) d_{1,\lambda}(k,q+k)e_0(q+k)]
\\
\hspace{-4mm}
&& \hspace{-6mm}
G^{\mu\nu\lambda\sigma}_7(p,r,s,q) =
   \int_k\tr[\gamma_5 e_0(p+r+s+k)d_{1,\mu}(p+r+s+k,r+s+k) \times \nn
&& \hspace{40mm}
     e_0(r+s+k)d_{1,\nu}(r+s+k,s+k)e_0(s+k) \times \nn
&& \hspace{40mm}
     d_{1,\lambda}(s+k,k)e_0(k)d_{1,\sigma}(k,q+k)e_0(q+k)]
\EAN
where (\ref{eq:d1pk}), (\ref{eq:d2pk}) and (\ref{eq:dnpk}) ( for $ n=3 $ )
have been used, and $ {\tilde A}_{\mu}^{(n)}(p) $, etc. are
defined in Eq. (\ref{eq:An}).

In the following, we show that all $ G's $ vanish in the continuum
limit except $ G_7 $, by expanding each one of them in power series
of $ p $, $ q $, $ r $ and $ s $, respectively.

First consider
\beq
G^{\mu\nu}_1(p,q) = G^{\mu\nu}_1(0,0) + O(a) \ ,
\eeq
where
\BAN
G^{\mu\nu}_1(0,0)
= \int_k\tr[\gamma_5 e_0(k)d_{1,\mu}(k,k)e_0(k)d_{3,\nu}(k,k)e_0(k)]
= 0 \ ,
\EAN
since its integrand vanishes,
\BAN
& & \tr[\gamma_5 e_0(k)d_{1,\mu}(k,k)e_0(k)d_{3,\nu}(k,k)e_0(k)] \nn
&=&  \tr(\Id)\epsilon_{\alpha\beta\gamma\delta}
     b(k)c^{\alpha}(k)c^{\gamma}(k)\partial_{\mu}d_0^{\beta}(k)
     d_{3,\nu}^{\delta}(k) = 0 \ ,
\EAN
where (\ref{eq:e0}), (\ref{eq:d1pp}), (\ref{eq:dnpk}), (\ref{eq:trace_0}),
(\ref{eq:trace_4}) and (\ref{eq:antisym}) have been used.
Similarly, we obtain
\BAN
G^{\mu\nu}_2 (0,0) = G^{\mu\nu}_3(0,0) = 0 \ .
\EAN

Next consider
\beq
G^{\mu\nu\lambda}_4(p,r,q) = G^{\mu\nu\lambda}_4(0,0,0) + O(a) \ ,
\eeq
where
\bea
& & G^{\mu\nu\lambda}_4 (0,0,0) \nn
&=&  \int_k\tr[\gamma_5 e_0(k)d_{1,\mu}(k,k)e_0(k)d_{1,\nu}(k,k)e_0(k)
    d_{2,\lambda}(k,k)e_0(k)] = 0 \ ,
\eea
since its integrand vanishes,
\bea
& &   \tr[\gamma_5 e_0(k) \partial_{\mu}e_0^{-1}(k)e_0(k)
	\partial_{\nu}d_0(k)e_0(k)d_{2,\lambda}(k,k)e_0(k)] \nn
&=&
      - \tr[\gamma_5 \partial_{\mu}e_0(k)
	\partial_{\nu}d_0(k)e_0(k)d_{2,\lambda}(k,k)e_0(k)] \nn
&=&
    - \tr(\Id) \Bigl\{\epsilon_{\alpha\beta\gamma\delta}
	\partial_{\mu}b(k)c^{\beta}(k)c^{\delta}(k)\partial_{\nu}
        d_0^{\alpha}(k){d}_{2,\lambda}^{\gamma}(k) \Bigr\} + \nn
& &
    + \tr(\Id) \Bigl\{ ( \epsilon_{\alpha\beta\gamma\delta}
                +\epsilon_{\alpha\beta\delta\gamma} )
	b(k)c^{\gamma}(k)\partial_{\mu}c^{\alpha}(k)
        \partial_{\nu}d_0^{\beta}(k){d}_{2,\lambda}^{\delta}(k)\Bigr\} \nn
&=& 0 \ ,
\eea
where (\ref{eq:e0}), (\ref{eq:d1pp}), (\ref{eq:eie}),
(\ref{eq:d2pk}), (\ref{eq:trace_0}), (\ref{eq:trace_4}) and
(\ref{eq:antisym}) have been used.
By the same token, we have
\bea
G^{\mu\nu\lambda}_5 (0,0,0) = G^{\mu\nu\lambda}_6 (0,0,0) = 0 \ .
\eea

Finally, we consider
\beq
\label{eq:G7}
G^{\mu\nu\lambda\sigma}_7(p,r,s,q) =
G^{\mu\nu\lambda\sigma}_7(0,0,0,0) + O(a)
\eeq
where
\bea
& & G^{\mu\nu\lambda\sigma}_7(0,0,0,0) \nn
&=& \int_k\tr[\gamma_5 e_0(k)d_{1,\mu}(k,k) e_0(k)d_{1,\nu}(k,k)
	e_0(k)d_{1,\lambda}(k,k) e_0(k)d_{1,\sigma}(k,k) e_0(k) ] \nn
&=& \int_k\tr[\gamma_5 e_0(k)\partial_{\mu}e_0^{-1}(k) e_0(k)
	\partial_{\nu}e_0^{-1}(k) e_0(k)\partial_{\lambda}e_0^{-1}(k) e_0(k)
	\partial_{\sigma}e_0^{-1}(k) e_0(k)] \nn
&=&
     \int_k\tr[\gamma_5 \partial_{\mu}e_0(k) \partial_{\nu}e_0^{-1}(k)
           e_0(k) \partial_{\lambda}e_0^{-1}(k) \partial_{\sigma}e_0(k) ] \ .
\label{eq:G7_0}
\eea
Here we have used (\ref{eq:d1pp}) and (\ref{eq:eie}) to simplify the
integrand. Comparing (\ref{eq:G7_0}) with (\ref{eq:G22_d}), one immediately
obtains
\bea
\label{eq:G7_00}
   G^{\mu\nu\lambda\sigma}_7 (0,0,0,0) = I_{\mu\nu\lambda\sigma}
= - \frac{N_f}{8\pi^2} \epsilon_{\mu\nu\lambda\sigma} \ ,
\eea
which has been evaluated in (\ref{eq:const}).

Thus, in the continuum limit, only $ G_7 $ has nonzero contribution
to the axial anomaly. Substituting (\ref{eq:G7}) into (\ref{eq:anx_4a}),
we obtain the axial anomaly at the order $ g^4 $,
\bea
\hspace{-2mm}
{\cal A}_4(x)
\hspace{-2mm}
&=&
\hspace{-2mm}
-\frac{g^4 N_f}{8\pi^2} \sum_{\mu\nu\lambda\sigma}
	\epsilon_{\mu\nu\lambda\sigma}
	\int_p\int_r\int_s\int_q e^{i(p+r+s-q)\cdot x}
	{\tilde A}_{\mu}(p){\tilde A}_{\nu}(r){\tilde A}_{\lambda}(s)
        {\tilde A}_{\sigma}(-q)  \nn
&=&    -\frac{g^4 N_f}{8\pi^2} \sum_{\mu\nu\lambda\sigma}
	\epsilon_{\mu\nu\lambda\sigma}
	\tr\bigl\{ A_{\mu}(x)A_{\nu}(x)A_{\lambda}(x)A_{\sigma}(x)\bigr\}
         \nn
&=&    -\frac{g^4 N_f}{32\pi^2} \sum_{\mu\nu\lambda\sigma}
	\epsilon_{\mu\nu\lambda\sigma}
	\tr\bigl\{ [A_{\mu}(x),A_{\nu}(x)]\,[A_{\lambda}(x),A_{\sigma}(x)]
        \bigr\} \ .
\label{eq:anx_g4}
\eea

Finally, adding (\ref{eq:anx_g2}), (\ref{eq:anx_g3}) and (\ref{eq:anx_g4})
together gives the axial anomaly (\ref{eq:anx}) in the continuum limit,
\bea
\label{eq:ax_an}
{\cal A} (x) &=& \tr[ \gamma_5 ( \Id - a r D )(x,x)] \nn
&=& \frac{g^2 N_f}{32 \pi^2} \sum_{\mu\nu\lambda\sigma}
                   \epsilon_{\mu\nu\lambda\sigma} \
          \tr \{ F_{\mu\nu}(x) F_{\lambda\sigma}(x) \} \ ,
\eea
where
\bea
F_{\mu\nu}=\partial_\mu A_\nu - \partial_\nu A_\mu + ig [A_\mu, A_\nu] \ .
\eea
Note that the axial anomaly (\ref{eq:ax_an}) is {\it invariant }
for any $ r > 0 $, since the anomaly coefficient (\ref{eq:const})
is independent of $ r $.

This completes our perturbation calculation of the axial anomaly
of the Ginsparg-Wilson lattice Dirac operator proposed in
Ref. \cite{Chiu:2001bg}.

\section{Concluding Remarks}

Several remarks are as follows.

It is obvious that (\ref{eq:ax_an}) also holds for other
$ T_\mu $ (\ref{eq:f}) such that $ D $ is doublers-free,
exponentially-local, and has correct continuum behavior.
In particular, (\ref{eq:ax_an}) holds for
\bea
\label{eq:f_alpha}
f = \left( \frac{2c}{\sqrt{t^2 + w^2} + w } \right)^{\alpha} \ ,
\hspace{4mm} \alpha \ge \frac{1}{2} \ , \hspace{4mm} 0 < c < 2 \ ,
\eea
as proposed in Ref. \cite{Chiu:2001bg}.

It is instructive to examine how well the continuum axial anomaly can be
recovered on a finite lattice, by evaluating the integral $ I_{1234} $
(\ref{eq:G22_e}) as a numerical sum over the discrete momenta
on a finite lattice. In Table 1, the ratios of $ I_{1234} $
to $ 1/8\pi^2 $ are listed for several
lattice sizes ( $ L^4 $ ), as well as for a range of $ r $, respectively.
Here the lattice spacing $ a $ is set to one, and the value of $ c $
[ see Eqs. (\ref{eq:f}) and (\ref{eq:w}) ] is fixed at 1.0.
Evidently, the integral $ I_{1234} $ tends to the continuum
value $ 1/8\pi^2 $ as $ L \to \infty $, independent of
the parameter $ r $.
This also provides a verification of our ansatz
to substitute $ d_0(K) $ by $ d_0^{-1}(k) $ in evaluating
$ I_{\mu\nu\lambda\sigma} $ (\ref{eq:G22_f}) in the continuum limit.

{\footnotesize
\begin{table}
\begin{center}
\begin{tabular}{|c|c|c|c|c|}
\hline
$ L $  & \multicolumn{4}{c|}{ $ r $ }  \\
\hline
     &   0.5  &  1.0   &   2.0  &  4.0    \\
\hline
\hline
%
  16  &     0.9204  &    0.9800  &   0.9848  &   0.7617   \\
\hline
  32  &     0.9810  &    0.9952  &   0.9988  &   0.9907   \\
\hline
  64  &     0.9953  &    0.9988  &   0.9997  &   0.9999   \\
\hline
  128 &     0.9988  &    0.9997  &   0.9999  &   1.0000   \\
\hline
  256 &     0.9997  &    0.9999  &   1.0000  &   1.0000   \\
\hline
\end{tabular}
\end{center}
\caption{
The ratio of the integral $ I_{1234} $ [ Eq. (\ref{eq:G22_e}) ] to
$ 1/8\pi^2 $,
for lattices of sizes $ 16^4, 32^4, 64^4, 128^4, 256^4 $, 
as well as for $ r = 0.5, 1.0, 2.0, 4.0 $, respectively.
The lattice spacing $ a $ is set to one,
and the parameter $ c $ [ Eqs. (\ref{eq:f}) and (\ref{eq:w}) ]
is fixed at $ 1.0 $. }
\label{table:1}
\end{table}
}

Next we repeat the same calculations of $ I_{1234} $ for another $ f $,
namely, replacing (\ref{eq:f}) by
\bea
\label{eq:f_1}
f = \frac{2c}{\sqrt{t^2 + w^2} + w } \ .
\eea
The results are listed in Table 2, which show that the integral
$ I_{1234} $ ( as a function of lattice size $ L^4 $ ) approaches
the continuum value $ 1/8\pi^2 $ much faster than that in Table 1.
This indicates that (\ref{eq:f_1}) may be a better choice than (\ref{eq:f}),
especially for small lattices.

In general, $ T_\mu $ (\ref{eq:Dc}) may not be in the form
$ T_\mu = f t_\mu f $,
then our present perturbation calculation may not go through without
modifications. We refer to our former derivation \cite{Chiu:1998qv}
for the general case, though some of our intermediate steps
need further clarifications. If one assumes that the coefficient of the
axial anomaly, $ g^2/32 \pi^2 $, is the same for $ U(1) $ and $ SU(n) $
background gauge fields, then it can also be determined \cite{Chiu:1999xf}
by imposing a gauge configuration with constant field tensors to
L\"uscher's formula \cite{Luscher:1999kn} for the axial anomaly of
Ginsparg-Wilson lattice Dirac fermions in a $ U(1) $ background gauge field,
provided that $ D $ is {\it topologically-proper}. However, for
a {\it topologically-trivial} Ginsparg-Wilson lattice Dirac operator
such as (\ref{eq:DcD}), it seems to be necessary to perform an explicit
calculation in order to determine its axial anomaly.
In passing, we also refer to other axial anomaly calculations
\cite{Kikukawa:1999pd,Adams:1998eg,Fujikawa:1999if,Suzuki:1999yz} for
the overlap Dirac operator \cite{Neuberger:1998fp,Narayanan:1995gw},
as well as that in the original Ginsparg-Wilson paper \cite{Ginsparg:1982bj}.
So far, a {\it nonperturbative analytic} calculation of the axial anomaly
for any Ginsparg-Wilson lattice Dirac operator, on a {\it finite} lattice,
is still lacking.

{\footnotesize
\begin{table}
\begin{center}
\begin{tabular}{|c|c|c|c|c|}
\hline
$ L $  & \multicolumn{4}{c|}{ $ r $ }  \\
\hline
     &   0.5  &  1.0   &   2.0  &  4.0    \\
\hline
\hline
%
  16  &     1.0000  &    1.0000  &   0.9911  &   0.7565   \\
\hline
  32  &     1.0000  &    1.0000  &   1.0000  &   0.9914   \\
\hline
  64  &     1.0000  &    1.0000  &   1.0000  &   1.0000   \\
\hline
  128 &     1.0000  &    1.0000  &   1.0000  &   1.0000   \\
\hline
  256 &     1.0000  &    1.0000  &   1.0000  &   1.0000   \\
\hline
\end{tabular}
\end{center}
\caption{
The ratio of the integral $ I_{1234} $ [ Eq. (\ref{eq:G22_e}) ] to
$ 1/8\pi^2 $, for $ f = 2c/( \sqrt{t^2 + w^2} + w ) $.
Other parameters are the same as those in Table 1. }
\label{table:2}
\end{table}
}

In summary, we have shown that the lattice Dirac operator (\ref{eq:DcD})
reproduces continuum axial anomaly for smooth gauge configurations,
even though it does {\it not} possess any topological
zero modes in topologically-nontrivial gauge backgrounds.
If one insists that the topologically zero modes of a lattice Dirac operator
are crucial for lattice QCD to reproduce the low energy hadron phenomenology,
then one should assure that a Ginsparg-Wilson lattice Dirac operator is
indeed {\it topologically-proper},
before it could be employed for lattice QCD computations.
However, so far, there does not seem to have compelling experimental
evidence that these topological zero modes are physically relevant,
unlike the axial anomaly in the trivial gauge sector,
which accounts for the decay rate of the neutral pion.
So there might be a very slight possibility that lattice QCD with
{\it topologically-trivial} quarks could reproduce low energy hadron
phenomenology. These issues seem to deserve further studies.

\appendix

\section{Basic Formulas in Weak Coupling Perturbation Theory}

In this appendix, we set up our notations by deriving some basic
formulas in weak coupling perturbation theory.
These formulas are used in our anomaly calculation in Section 2.

Consider the forward and backward difference operators
\bea
\nabla^{+}_{\mu}(x,y) &\equiv& U_{\mu}(x)\delta_{x+a\hat\mu,y}
                                -\delta_{x,y} \ , \\
\nabla^{-}_{\mu}(x,y) &\equiv& \delta_{x,y}-
                U^{\dagger}_{\mu}(x-a \hat\mu)\delta_{x-a\hat\mu,y} \ ,
\eea
where the link variables ( for $ SU(n) $ gauge group ) are defined as
\bea
U_\mu(x) &=& \exp
 \left[ i a g A_\mu \left(x+ \frac{a}{2} \hat\mu \right) \right] \ , \\
U_{\mu}^{\dagger} (x-a\hat\mu) &=& \exp
 \left[ - iag A_{\mu} \left(x- \frac{a}{2} \hat\mu \right) \right] \ .
\eea
Expanding $ \nabla^{\pm} $ in power series of the gauge coupling $g$,
\bea
\nabla^{\pm}_{\mu}={(\nabla^{\pm}_{\mu})}_0 + {(\nabla^{\pm}_{\mu})}_1
	+ {(\nabla^{\pm}_{\mu})}_2 + \cdots
\eea
and performing the Fourier transform
\beq
\sum_{x,y} e^{-i(p\cdot x - q\cdot y)} {(\nabla^{+}_{\mu})}_n (x,y)
\equiv {(\nabla^{+}_{\mu})}_n(p,q) \ ,
\eeq
we obtain
\beq
{(\nabla^{+}_{\mu})}_0(p,q) = (e^{ip_{\mu}a}-1) \ \delta^4(p-q)
\equiv {(\nabla^{+}_{\mu})}_0(p) \ \delta^4(p-q)
\eeq
and
\bea
{(\nabla^{+}_{\mu})}_n(p,q) &=&
     \sum_{x,y}e^{-i(p\cdot x-q\cdot y)}{1\over n!} (iag)^n
 \left[ A_{\mu} \left(x+\frac{a}{2} \hat\mu \right) \right]^n
 \delta_{x + a \hat\mu,y} \nn
&=&
\sum_x e^{-i(p-q)\cdot(x+ a \hat\mu/2)}{1\over n!} (iag)^n
 \left[ A_{\mu} \left(x+\frac{a}{2} \hat\mu \right) \right]^n
 e^{i(p_{\mu}+q_{\mu})a/2} \nn
&=&
     \frac{(iag)^n}{n!} {\tilde A}^{(n)}_{\mu}(p-q)
     e^{i(p_{\mu}+q_{\mu})a/2} \nn
&=&
     \frac{g^n}{n!} {\tilde A}^{(n)}_{\mu}(p-q)
     \partial_{\mu}^n {(\nabla^+_{\mu})}_0 \LL({p+ q \over 2}\RR)
\eea
where
\bea
\label{eq:An}
 {\tilde A}^{(n)}_{\mu}(p-q) &=& \sum_x e^{-i(p-q)\cdot(x+ a \hat\mu/2)}
  \left[ A_{\mu} \left(x+\frac{a}{2} \hat\mu \right) \right]^n \ , \\
 {(\nabla^+_{\mu})}_0 \LL({p+q \over 2}\RR) &=& e^{i(p_\mu + q_\mu )a/2}-1 \ .
\eea

Similarly, we have
\bea
{(\nabla^{-}_{\mu})}_0(p,q) &=& (1 - e^{-ip_{\mu}a}) \ \delta^4(p-q)
\equiv {(\nabla^{-}_{\mu})}_0(p) \ \delta^4(p-q) \ , \\
{(\nabla^{-}_{\mu})}_n(p,q) &=&
        {g^n\over n!} {\tilde A}^{(n)}_{\mu}(p-q)
        \partial_{\mu}^{n} {(\nabla^-_{\mu})}_0 \LL({p+q\over 2}\RR) \ .
\eea

Then we obtain the weak coupling perturbation series of
$ t^{\mu} $ (\ref{eq:tmu}) and $ w $ (\ref{eq:w}) in the momentum space :

\bea
t^{\mu}(p,q) &=& {1\over 2}[\nabla_{\mu}^+(p,q)+\nabla_{\mu}^-(p,q)] \nn
  &=& t^{\mu}_0(p)\delta^4 (p-q) + t^{\mu}_1(p,q) + t^{\mu}_2(p,q) +\cdots \\
w(p,q) &=& c -\frac{1}{2}
              \sum_{\mu} [\nabla_{\mu}^+(p,q)+\nabla_{\mu}^-(p,q)] \nn
       &=& w_0(p) \delta^4 (p-q) + w_1(p,q) + w_2(p,q) + \cdots
\eea
where
\bea
t^{\mu}_0(p) &=& i \sin(p_{\mu}a) \ , \\
w_0(p)       &=& c-\sum_{\mu}[1-\cos(p_{\mu}a)] \ , \\
\label{eq:tmun}
t^{\mu}_n(p,q) &=& \frac{g^n}{n!} {\tilde A}^{(n)}_{\mu}(p-q)
                   \partial^n_{\mu} t^{\mu}_0\LL(p+q\over 2\RR) \ , \\
\label{eq:wn}
w_n(p,q) &=& \frac{g^n}{n!}
        \sum_{\mu} {\tilde A}^{(n)}_{\mu}(p-q) \partial^n_{\mu}
        w_0\LL(p+q\over 2\RR) \ .
\eea

\eject

\bigskip
\bigskip
\flushpar
{\bf Acknowledgement }
\bigskip

\noindent

This work was supported in part by the National Science Council,
Republic of China, under the grant numbers NSC89-2112-M002-079,
and NSC90-2112-M002-021.
We would like to thank David Adams for discussions.

\bigskip
\bigskip


\end{document}